\documentstyle[11pt,newpasp,twoside,epsf]{article}
\def\edcomment#1{\iffalse\marginpar{\raggedright\sl#1\/}\else\relax\fi}
\reversemarginpar

\begin{document}
\title{The Be star content of young open clusters}
 \author{Juan Fabregat}
\affil{Observatorio Astron\'omico, Universidad de Valencia, 46100
Burjassot, Spain}
\affil{GEPI/FRE K2459 du CNRS, Observatoire de Paris, 92195 Meudon,
France}

\begin{abstract}
We present a photometric survey aimed to characterize the Be star
population of young open clusters. It is found that in these clusters
early-type Be stars are more frequent than in the galactic field, and
late-type Be stars are scarce or inexistent. We interpret this result as
evidence for an evolutionary enhancement of the Be phenomenon towards
the end of the main sequence lifetime.
\end{abstract}

\section{Introduction}

The study of the frequencies of Be stars in open clusters of different
ages and their distribution as a function of the spectral subtype are key
issues in the discussion on the evolutionary status of Be stars. 

The clusters NGC 663, NGC 869 and NGC 884 are known to be among the
galactic clusters with a higher Be star content. Lists of Be stars in
these clusters with information on spectral types are presented by
Mermilliod (1982) and Slettebak (1985). From these lists it is remarkable
that all Be stars are of early spectral types: earlier than B3 in NGC 869
and 884 and earlier than B5 in NGC 663. This fact contrasts with the
distribution of the galactic Be star population. About 20\% of B stars are
Be, along the B0-B8 range (Zorec \& Briot 1997).

A possible explanation of this discrepancy is observational bias. Most of
the known Be stars in open clusters have been identified within the
framework of large surveys for emission line stars in the Milky Way, which
are magnitude limited. For most clusters only the brightest stars have
been searched for line emission, and hence the derived frequencies are
only lower limits, restricted to the earliest spectral types.

The aim of this work is to study the complete B star sequence of the
young open clusters with a high Be star content referred to above, and
to analyze the Be star frequency for the different spectral subtypes.
To detect emission line stars we have used CCD imaging photometry through
interference filters centered in the H$\alpha$ and H$\beta$ Balmer lines.

\section{Observations and reduction procedure}
 
Observations were done during two runs in December 1995 and November
1998 at the Calar Alto Observatory (Almer\'\i a, Spain). The first run was
performed with the 1.23m telescope of the Centro Astron\'omico Hispano 
Alem\'an, and the second one with the 1.52m. telescope of the
Observatorio Astron\'omico Nacional.
 
In each telescope we used two pairs of interference filters centered on
H$\alpha$ and H$\beta$ Balmer lines. Each pair consisted of a wide and a
narrow filter.
 
The data have been reduced using the standard procedures in IRAF. For
each frame we have obtained the instrumental magnitudes of the stars. We
have transformed the instrumental magnitudes into the $\alpha$ and $\beta$ 
indices by subtracting the magnitude in the wide filter from the
magnitude in the narrow one.   

\begin{figure} 
\plotone{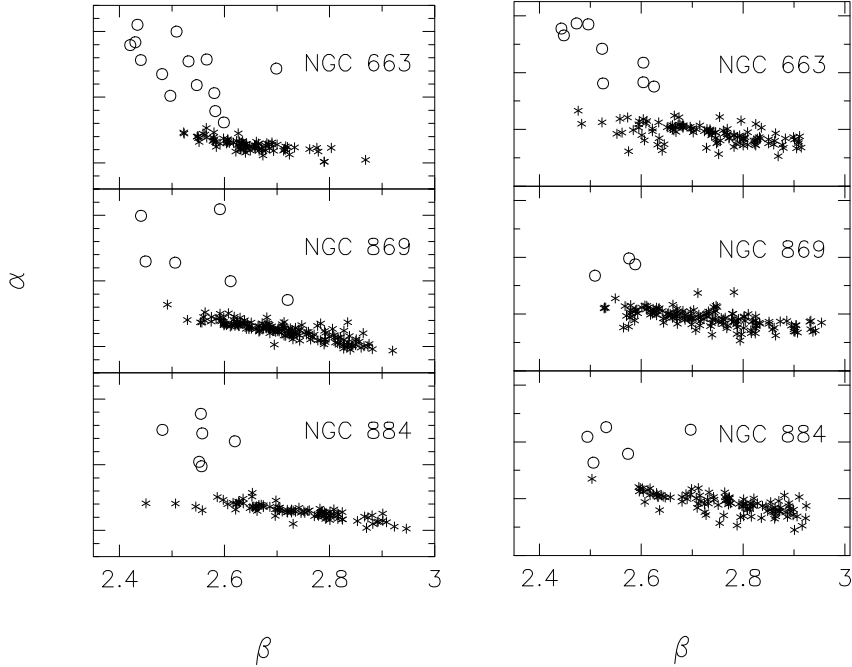}
\caption{$\alpha - \beta$ photometric diagram of the observed clusters. 
Left and right panels correspond to 1995 and 1998 data, respectively. Be 
stars are represented with open circles} 
\end{figure}

We have transformed our instrumental $\beta$ indices into the standard
Crawford \& Mander (1966) system. As there is not a widely
accepted H$\alpha$ photometric system, we have worked with the
instrumental indices obtained in our observations. Note that the
instrumental $\alpha$ systems are different in the two runs, which explain
the different scales in Figure 1.
 
There is a good correlation between the $\alpha$ and $\beta$ indices for B
stars, and hence in the plot of the two indices for objects without
emission a well-defined sequence is apparent. For emission-line stars,
since emission in H$\alpha$  is much stronger than in H$\beta$, such
objects appear significantly above the sequence. In Figure 1 we have
represented the photometric $\alpha - \beta$ plane for the observed
clusters in two epochs. Be stars are represented with a different symbol.
   
In the 1998 run Str\"omgren $uvby$ photometry was also obtained, and has
been published elsewhere (Capilla \& Fabregat 2002a,b). The $uvby$
photometric diagrams show that our survey is complete through all the
B-type main sequence in the three observed clusters.

\section{Discussion}

From the $uvby$ and H$\beta$ photometry we have obtained the reddening and
intrinsic colours and indices for the stars observed (see Capilla \&
Fabregat 2002a,b for details). Photometric spectral types were derived
from the $c_0$ index, by using the mean $c_0 - $ spectral-type relation
presented in Table II in Crawford (1978). For Be stars, of which the  
$c_0$ index is anomalous due to circumstellar emission in the Balmer 
continuum, spectral  types have been assigned following the precepts in 
Fabregat \& Torrej\'on (1998). The photometric spectral types derived in 
this way are coincident with spectroscopic MK types presented by Slettebak 
(1985) for the stars in common, within one spectral subtype in all cases.

From the $\alpha$ and $\beta$ photometry we have detected as emission line
stars most of the previously known Be stars, and we have made three new
detections. All but one are of spectral types earlier than B5. In NGC 869
and 884 we have obtained photometry for 168 stars in the range B4-A0, and
none of them presented line emission. In NGC 663 we observed 132 stars in
the interval B5-A0, and only one Be star was found (star NGC 663 51, of
spectral type B6). This proves that the lack of late-type Be stars in the
surveyed clusters is a real feature and not observational bias.

\begin{figure} 
\plotone{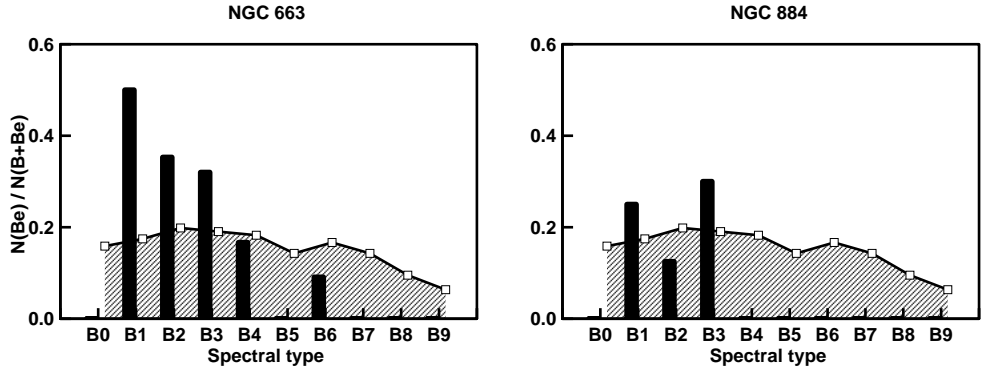}
\caption{Relation between Be star frequencies and spectral subtypes. Bars 
represent the observed frequencies in the clusters. The solid line and 
shadowed area are the mean frequencies for field stars in the Galaxy.} 
\end{figure}

In Figure 2 we present the histogram of the Be star frequencies for NGC 
663 and 884. NGC 869 was not analyzed due to the small number of Be stars
in the surveyed area. The solid line and shadowed area in Fig.~2
represent the frequencies for the galactic field population, as given by
Zorec \& Briot (1997). It is apparent that the frequencies in the open
clusters and in the field population are significantly different. In both
clusters, the frequency of early-type Be stars is similar or greater than
in the galactic field. Conversely, both clusters almost completely lack
late-type Be stars, which have a frequency of about 20\% in the galactic
field.

In order to check the validity of this result for a larger cluster sample, 
we have analyzed in a similar way four open clusters in the Magellanic 
Clouds (hereafter referred to as MC). We have used the data presented by 
Keller, Bessell, \& Da Costa (2000). Spectral types have been derived from 
the V magnitudes given by these authors. 

\begin{figure} 
\plotone{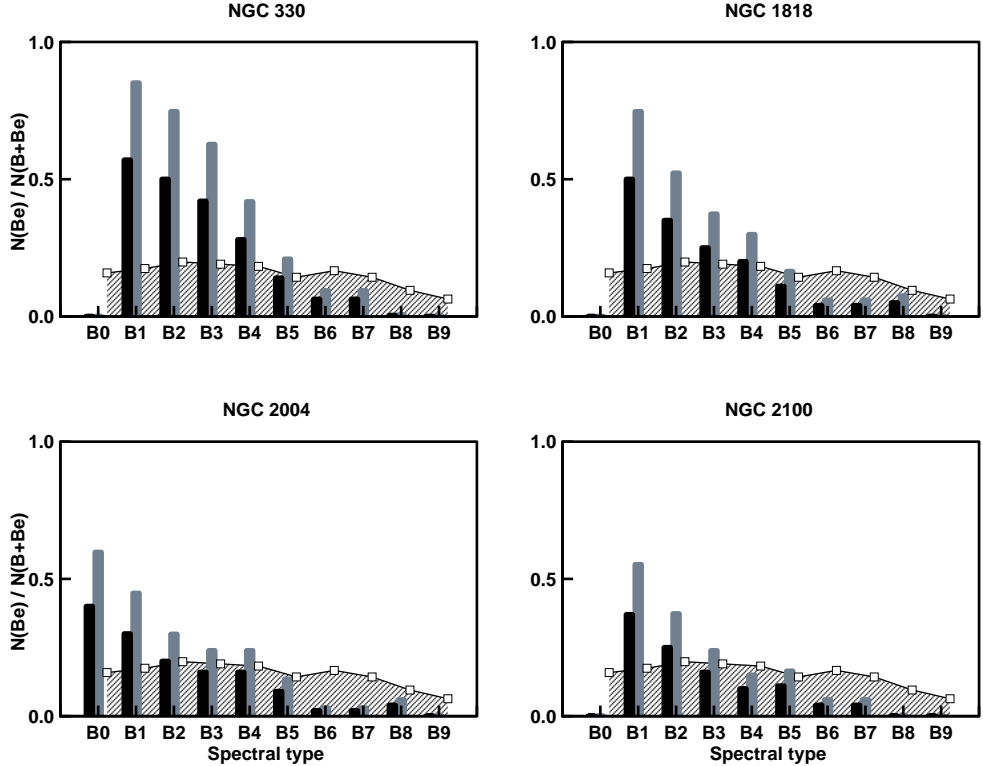}
\caption{Relation between Be star frequencies and spectral subtypes for
clusters in the Magellanic Clouds. Black bars represent the observed
frequencies in the clusters, and grey  bars frequencies corrected by the
expected detection efficiency (see text). Solid line and shadowed area as
in Fig.~2.} 
\end{figure} 

The Be star frequencies in Keller et al. (2000) have been derived from a 
single epoch photometric survey. It is well known that these surveys lose
a significant fraction of the actual number of Be stars, due to the
episodic lost of emission lines in many Be stars and the inability of the
photometric technique to detect faint emitters. Conversely, the study on
NGC 663 and 884 is based on all Be stars detected in more than 80 years of
spectroscopic and photometric surveys. To make both samples
comparable, we have estimated the ratio between the actual number of Be
stars and the number detected in a single epoch photometric survey. We
have considered as complete samples all the known Be stars in NGC 663, 869
and 884, and we have used the detections in our 1995 and 1998 data, and 
the observations by Goderya \& Schmidt (1994) as single epoch surveys. We 
obtained that the ratio between the stars detected and the actual content
is around two thirds.

Results are presented in Figure 3. Each subtype bin is represented by two
bars in the histogram. The shorter one is the Be star frequency detected
by Keller et al. (2000), and the larger one is this number corrected by
the detection efficiency parameter derived in the above
paragraph.  As in the galactic clusters, the frequency of early-type Be
stars in the MC clusters is much higher than in the Milky Way field, while
late-type Be stars are significantly less abundant. 

We conclude that the Be star population in young open clusters is
significantly different, and hence not representative, of the mean
population in the Galaxy. We interpret this fact in the context of the
evolutionary hypothesis for the Be phenomenon proposed by
Fabregat \& Torrej\'on (2000), in which the Be star phase would appear
during the second half of the main sequence lifetime of a B star. In the
age range of the analyzed clusters, early B stars are close to the end of
their main sequence lifetime, and hence the Be phenomenon is much more
frequent than in a mixed age field sample. Late-type B stars are at
the beginning of the main sequence, where the Be phase would be scarce or
inexistent.

\begin{figure} 
\plotone{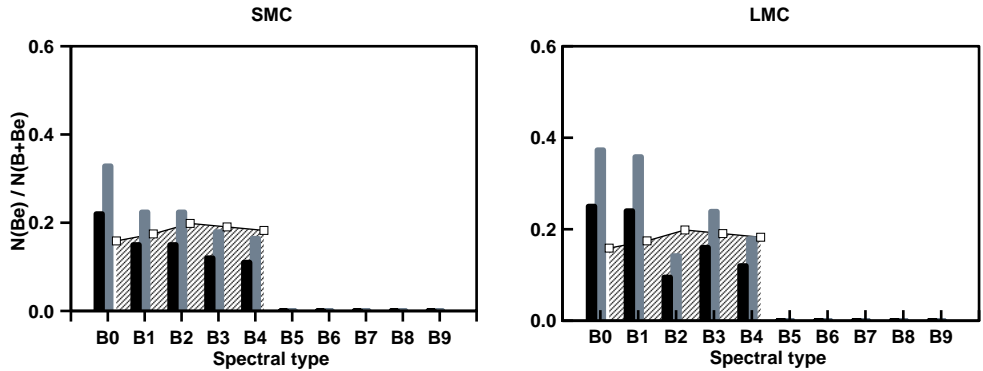}
\caption{Relation between Be star frequencies and spectral subtypes for MC 
fields. Bars, solid line and shadowed area as in Fig.~3.} 
\end{figure} 

Maeder, Grebel \& Mermilliod (1999) have recently proposed that the Be
phenomenon is related to the metal abundance, being enhanced in a low
metallicity environment. They based their conclusions on the study of the 
Be frequencies in several young clusters in the inner and outer Galaxy 
and the MC. The fact that the Be population in young clusters is not
representative of their host galaxy casts doubts on this conclusion.
In order to ascertain this issue we have also compared the Be
star frequencies in the LMC and SMC fields with those of the Galaxy
field. Result are presented in Figure 4. 

The data used are from Keller, Wood and Bessell (1999). 
Their survey of several MC fields around open clusters only reach the 
early B range, and the comparison has to be
restricted to this range. It is apparent, however, that the frequencies in
the MC fields are compatible, within errors, with those of the Milky Way
field. To ascertain the reality of the metallicity dependence of the Be
phenomenon a more complete survey of the LMC and SMC field population has
to be done.

\section{Conclusions}

We have shown that the Be star content of young open clusters is
significantly different that the content of the mean galactic
population. This is an evidence for the evolutionary enhancement of the Be
phenomenon during the second half of the main sequence lifetime.

\end{document}